\documentclass{article}
\usepackage{graphicx} 

\title{PBH Paper}
\author{Paul Halpern}
\date{November 2025}

\begin{document}

\fontsize{10 pt}{13 pt}
\selectfont
\begin{center}
{\bf PROBING FOR PRIMORDIAL BLACK HOLE CANDIDATES IN EXOPLANET SEARCH DATA} \vspace{1 cm}
\fontsize{8pt}{10pt} \selectfont\\
\vspace{2 pt}
{\it Paul Halpern$^{\ast}$, Erik Cauley, Max Stoltzmann, and Mauritz Wilshusen\\Department of Physics,\\
Saint Joseph's University, 5600 City Avenue,\\
Philadelphia, PA. 19131, USA\\$^{\ast}$phalpern@sju.edu}
\end{center}
\begin{quote}
\fontsize{8pt}{10pt} \selectfont We have sifted through astrophysical data collected by various radial velocity and gravitational microlensing searches for exoplanets with the goal of identifying potential signs of the presence of primordial black holes (PBH). Our motivation is that those hypothesized remnants of inhomogeneous energetic fluctuations in the early universe, though too small for direct detection, are thought to have a mass range similar to that of planets. Thereby, if captured by stars, they could conceivably make their presence known through stellar wobbles picked up by means of Doppler spectroscopy in the radial velocity method, or alternatively through microlensing. In our analysis of such data, we have identified potential PBH contenders by ruling out any exoplanet candidates that have been detected through direct imaging or transit methods, as they would have sizes consistent with planets rather than PBHs. In particular we focus on the objects Kepler-21 Ac, HD 219134 f, Gliese 686 b, HR 5183 b, HD 20794 e, and Wolf 1061 d, each of which has been found using the radial velocity method but never imaged (either directly or through transit). We also examine the microlensing events MOA 2009-BLG-387L and OGLE-2016-BLG-1540, which offer promise as candidate PBHs. We present these as a representative, but not exclusive list, of potential PBH contenders. Furthermore, future imaging, especially focused on signals of planetary dimensions versus evaporation signatures, might clarify which of these are indeed exoplanets.


\end{quote}
\fontsize{10 pt}{13 pt} \selectfont \vspace{1 cm} \noindent
 {\bf 1. Introduction}

\vspace{6 pt} \noindent Over the past three decades various methods have successfully been used to detect exoplanet candidates, including the radial velocity method, microlensing, astrometry, and various types of transit methods and direct imaging techniques.  While transit, involving looking for periodic dips in the star's brightness as the planet passes in front of its line of sight, is by far the most fruitful method, representing more than 4000 candidates to date, the radial velocity technique has also been productive, revealing more than 1000 candidates. \cite{ortiz}

Because of the limits of each, the various discovery methods work best for exoplanet identification when taken in tandem.  For example, while the radial velocity technique offers a range of masses for each candidate, as well as orbital periods, it does not directly offer information about size.  If transit or direct imaging successfully targets the same object, then size information emerges through such further observations resulting in the absolute indication of an exoplanet of particular dimensions, Conversely, if, on the other hand, transit or direct imaging are applied to a candidate found through the radial velocity method, and no positive results emerge, it is conceivable that the radial velocity results might be subject to a different interpretation, such as relic bodies remaining from the early cosmos.

A topic of considerable interest for almost six decades has been the question of whether or not regions of excess density in the hot, turbulent primeval universe collapsed into massive compact objects, and any of their remnants are extant. \cite{zeldovich} Such hypothetical massive but minuscule entities are known as primordial black holes (PBH). \cite{novikov} Presuming they exist, such objects would comprise a laboratory for understanding the Hawking radiation process, proposed by Stephen Hawking in 1974 \cite{hawking}, by offering characteristic evaporation signatures.

The Hawking radiation temperature of a PBH is given by the standard formula:

\begin{equation}
T_H=\frac{\hbar c^3}{8\mathit{\pi G}k_bM}
\end{equation}

Over the billions of years since the nascent epoch of the universe, Hawking radiation would effectively decrease the mass and rotational energy of each PBH.  Depending on their initial mass, by the present era, some would vanish altogether, while others would remain and conceivably be detectable, at least indirectly. Whether or not such PBHs remain or have disappeared depends respectively on whether or not their lifetime is longer than the present age of the universe. 

The formula describing the lifetime of a black hole in seconds, is given by

\begin{equation}
t=\frac{10240\pi ^2G^2}{hc^4}M^3=\left(8.407x10^{-17}\frac{\mathit{s}}{\mathit{kg}^3}\right)M^3
\end{equation}

That effectively means that extant PBHs need to be above $10^{12}$ kg in mass, with a likely range between $10^{13}$ and $10^{22}$ kg.

Carr, K{\"u}hnel, and Sandstat have investigated the notion that PBHs constitute a credible candidate for dark matter. \cite{carr} Such unseen constituents would likely represent objects of planetary mass or smaller, invisible but abundant. Possibly, the merger of two such objects, if massive enough, might generate a characteristic and potentially detectable gravitational wave signature.

Given the prospect of an abundance of PBHs, researchers have explored the idea that stellar bodies might capture PBHs, which in turn would have a gravitational influence on the host star and orbiting planets. In our solar system there has been speculation of a hitherto unseen "Planet X" that affects the orbits of objects in its outer region. Rather than an actual planet, however, Jakub Scholtz and James Unwin have speculated that it could be a PBH instead. \cite{scholtz} Such capturing of PBHs would be rare, researchers Benjamin V. Lehmann, Ava Webber, Olivia G. Ross, and Stefano Profumo have pointed out, unless PBHs comprise the bulk of the dark matter in the universe \cite{lehmann} Nevertheless, as long as the prospect of their capture by stars remains an open question, it is interesting to consider other stellar systems and ponder if PBHs in those might be mistaken for extrasolar planets due to similar mass and gravitational pull.

\fontsize{10 pt}{13 pt} \selectfont \vspace{1 cm} \noindent
 {\bf 2. Searching for Primordial Black Holes in Radial Velocity Findings}

\vspace{6 pt} \noindent In the radial velocity technique, Doppler spectroscopy identifies stars that exhibit periodic "wobbles" in their motion with components along the line of sight due to the gravitational pull of objects such as binary companions or exoplanets. It has proven a reliable method of determining the mass range of bodies orbiting such stars.
Notably, the semi-amplitude (denoted by K) of the radial velocity variations in the spectrum of such stars, over the course of an orbital period P, is related to the mass function f of the gravitationally influencing object by:

\begin{equation}
f=\frac{PK^3\left(1-e^2\right)^{3/2}}{2\mathit{\pi G}}=\frac{M_{\mathit{unseen}}^3\sin
^3i}{\left(M_{\mathit{unseen}}+M_{\mathit{seen}}\right)^2}
\end{equation}
The motion of the unseen object is deduced from the seen object and the parameters are identical except K is scaled by a
factor of  $M_{\mathit{unseen}}/M_{\mathit{seen}}$ \ and the orientation of the elliptical orbit with respect to the
line of sight, denoted by  $\omega $, differs by  $\pi $. The application of radial velocities is not just limited to
stars, it can be used to detect unseen objects ranging in mass from planets to black holes. For exoplanets, the focus of
this study, the mass function can be approximated in the large-mass ratio limit 
$M_{\mathit{unseen}}/M_{\mathit{seen}}=M_{\mathit{planet}}/M_{\ast }{\ll}1$. What this limit does is produce an
equation relating the amplitude of a Doppler shift to the mass, inclination, and eccentricity of the planet:

\begin{equation}
K{\approx}\left(\frac{2\mathit{\pi G}}{PM_{\ast }^2}\right)^{\frac 1
3}\frac{M_{\mathit{planet}}\mathrm{sin}i}{\sqrt{1-e^2}}
\end{equation}

It is interesting to sift through the list of exoplanet candidates found by this method, identify which haven't yet been observed through transit or imaging and ponder if any of those offer promise as potential PBH contenders. We have identified six such radial velocity finds that might be ripe for further examination as possible PBHs. However our list is far from exhaustive, and might well change as new data comes in.

One of the most promising contenders for a potential PBH is the exoplanet candidate Kepler-21 Ac found by the radial velocity method in eccentric orbit around the subgiant star Kepler-21. It is a long-period super-Jupiter candidate that fits the general criteria for a PBH. During the Kepler mission in 2010 \cite{borucki}, the system was imaged, and Kepler-21 Ac could not be found transiting the host star. Yet Doppler spectroscopy data suggest that there is an object of $4$ Jupiter masses acting on Kepler 21. With a mass of $7.59×10^{27}$ kg the potential PBH sits in the credible mass range. Its calculated lifetime of $3.68×10^{68}$ years would far exceed the age of the universe, meaning that it would not yet have vanished due to Hawking radiation. If it were a PBH, its Schwarzschild radius would be only $11.28$ m, explaining the lack of imaging. Perhaps, if so, its evaporation signature might potentially be detected by a future probe of that region.

Another potential contender is HD 219134 f, thought to be a super-Earth object of mass $ 7.3 M_{\oplus}$ orbiting a K-type star with a period of 22.7 days. It was identified by the radial velocity method in 2015. Similarly, it has yet to be imaged or found via transit. If it were a PBH instead, that object would have a Schwarzschild radius of only $0.065 m$, well-explained its lack of detection through imaging or transit.

Gliese 686 b, thought to be a Neptune-like exoplanet candidate orbiting a M-type star with a period of 15.5 days, and HR 5183 b, a Jupiter-like object orbiting a G-type star with a period of 74 years, have each similarly failed to appear in imaging or transit data, despite being found through the radial velocity method. If they turned out to be PBHs instead, their calculated Schwarzschild radii of $0.0132$ m and $8.44$ m, respectively, would readily explain their lack of detection via directly imaging or transit surveys.

Finally, we consider HD 20794 e, a super-Earth orbiting a G-type star with a period of 147 days, and Wolf 1061 d, another super-Earth orbiting a red dwarf star with a period of 217 days.  Each has a highly eccentric orbit, suggesting the possibility of captured PBHs rather than exoplanets. Further research would be needed to distinguish those options, albeit with exoplanets remaining the most probable outcome.

\vspace{1 cm} \noindent
 {\bf 3. Identifying Primordial Black Hole Candidates in Microlensing Results}

\vspace{6 pt} \noindent Another potential avenue of exploration for potential PBH objects captured by stars is analyzing microlensing data. During this process, the object to be measured is treated as a gravitational lens. The background source is often seen as a blurry unresolved point. Over the course of the lensing event, light bending might distort or magnify the source in such a way that it gets dimmer and bright again in the following fashion:

\begin{equation}
m(t)=m_{\mathrm{base}} + s\,(t-t_0) + o - 2.5\log_{10}\!\left(A(t)\right)
\end{equation}

where:

\begin{equation}
A(t)=\frac{u(t)^2+2}{u(t)\sqrt{u(t)^2+4}},
\end{equation}

and:

\begin{equation}
u(t)=\sqrt{u_0^2 + \left(\frac{t-t_0}{t_E}\right)^2}.
\end{equation}

The angular Einstein radius can be defined as:

\begin{equation}
\theta_E=\sqrt{\frac{4GM}{c^2}\cdot\frac{D_{LS}}{D_L D_S}}.
\end{equation}

Hence, through the light curve of the lensed star it possible to determine the mass of the unseen object via the equation:

\begin{equation}
M=\frac{c^2}{4G}\cdot\frac{D_L D_{LS}}{D_S}\cdot(t_E v_T)^2.
\end{equation}

Therefore, it is possible to estimate the mass of such a lensing object by analyzing the brightness curve of an unresolved background source. If future investigation fails to reveal a transit or image, then it is possible that the lensing body represents a PBH. As previously discussed, the ideal PBH mass window is approximately $10^{24}\,$g to $10^{25}\,$g.  A very small Einstein radius increases the likelihood that a detected ``exoplanet'' may in fact be a PBH.

We downloaded raw data from the NASA Bulk Download site and developed a Python script that computes the masses and Einstein radii of 248 microlensing candidates. 
One of the most interesting candidates from this dataset is MOA~2009-BLG-387L.

\begin{figure}
    \centering
    \includegraphics[width=\linewidth]{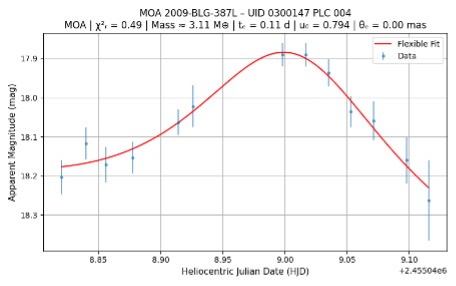}
    \caption{MOA~2009-BLG-387L}
    \label{fig:fig1}
\end{figure}

MOA-2009-BLG-387L is a gravitational microlensing event identified in the MOA survey, where the primary lens was estimated to have a mass of approximately $3.11\,M_{\oplus}$. 
The most striking feature is its extremely small Einstein angular radius, estimated at just $0.003$ milliarcseconds. 
This is significantly smaller than typical planetary or stellar lenses and implies either a highly compact lens or a distant, ultra-low-mass configuration. 
The lens is theoretically interpreted as a low-mass host star. 
Direct imaging in the future could either support the PBH hypothesis or rule out this candidate.

\begin{center}
\begin{tabular}{l l l}
\textbf{Aspect} & \textbf{Value / Observation} & \textbf{PBH Relevance} \\ \hline
Detection Method & Microlensing (MOA + OGLE) & gravitational effect only \\
Mass Estimate & $\sim 3.1\,M_{\oplus}$ & within PBH range $(10^{24}$--$10^{26}\,$g) \\
Einstein Radius & $\sim 0.003$ mas & extremely small, suggests compact body \\
Host Star Detected? & none & consistent with dark PBH lens \\
Lensing Events & two consistent detections & high reliability \\
Emission Signature & none observed & PBHs are non-accreting \\
Transit / Radius & none & PBHs do not transit \\
Formation Model & capture plausible & consistent with PBH scenario \\
\end{tabular}
\end{center}

Another interesting candidate, OGLE-2016-BLG-1540, can be found in the OGLE online dataset. It  was previously described in the literature as difficult to classify but optically undetectable.

\begin{figure}
    \centering
    \includegraphics[width=\linewidth]{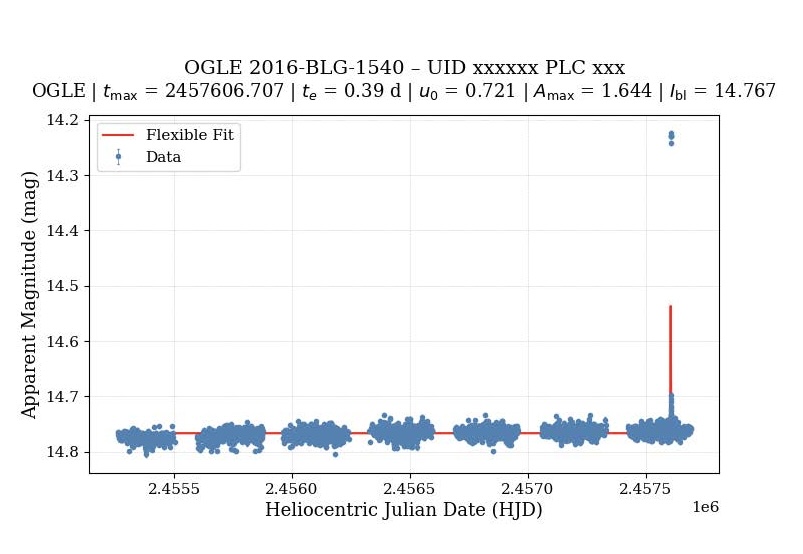}
    \caption{OGLE-2016-BLG-1540}
    \label{fig:fig2}
\end{figure}

OGLE-2016-BLG-1540 is a short-timescale microlensing event independently detected by OGLE-IV and KMTNet.  It exhibits a brief, isolated magnification without recurring signals, and is one of the shortest microlensing events with well-characterized finite-source effects. No host flux and no lens flux were detected, leading to the hypothesis that the lens is a dark, dense object. 
Its compact Einstein radius further strengthens the possibility of a PBH.

Given:
\[
t_E = 0.320 \pm 0.003~\mathrm{days}, \qquad
\theta_E = 9.2 \pm 0.5~\mu\mathrm{as} = 9.2\times 10^{-6}~\mathrm{arcsec},
\]
\[
\mu_{\mathrm{rel}} = 10.5~\mathrm{mas/yr}, \qquad
D_S = 8~\mathrm{kpc}, 
\]
with lens either in the disk $(D_L \approx 2~\mathrm{kpc})$ or bulge $(D_L \approx 6~\mathrm{kpc})$.

\vspace{6 pt} \noindent

The lens mass is:
\begin{equation}
M=\frac{\theta_E^2 c^2}{4G}\cdot\frac{D_L D_S}{D_S - D_L}.
\end{equation}

The angular Einstein radius (in radians), $\theta_E$ is:
\[
\theta_E = 9.2 \times 10^{-6}\times\frac{\pi}{648000}
\approx 4.46\times 10^{-11}~\mathrm{rad}.
\]

In the disk case: $D_L \approx 2 ~ \mathrm{kpc}$, therefore
$M \approx 2.1\times10^{25}~\mathrm{kg}
\approx 3.5\,M_{\oplus}$

\vspace{6 pt} \noindent

In the bulge case, on the other hand: $D_L \approx 6 ~ \mathrm{kpc}$, therefore

$M \approx 7.0\times10^{25}~\mathrm{kg}
\approx 11.7\,M_{\oplus}$

\vspace{6 pt} \noindent

Thus the mass of the lensing object lies between $3.5\,M_{\oplus}$ and $11.7\,M_{\oplus}$, a range well within the planetary-mass PBH window.

\vspace{1 cm} \noindent
 {\bf 4. Conclusion}

\vspace{6 pt} \noindent We have examined several exoplanet candidate events from the radial velocity and microlensing events that have yet to be imaged or further detected through transit. Promising candidates include Kepler-21 Ac, HD 219134 f, Gliese 686 b, HR 5183 b, HD 20794 e, and Wolf 1061 d, each found using the radial velocity method, and MOA-2009-BLG-387L and OGLE-2016-BLG-1540, each found using the microlensing method (a list that is far from exhaustive). Therefore we speculate that those objects might potentially constitute PBH objects. In each case, however, confirmation would arise only with the finding of an evaporation signatures or other evidence of orbiting PBHs. On the other hand, further studies using the transit or direct imaging method might indeed validate the more likely hypothesis of those objects representing exoplanets. Future work in this area should prove revealing.

\vspace{1 cm} \noindent
 {\bf Acknowledgements}

\vspace{6 pt} \noindent We wish to thank Saint Joseph's University for Summer Scholars awards that aided in the completion of this project.

\newpage

\vspace{1 cm} \noindent\fontsize{9 pt}{11 pt} \selectfont

\end{document}